# Experimental and numerical studies on kV scattered x-ray imaging for real-time image guidance in radiation therapy

Yanqi Huang[1,2,*], Kai Yang[3,*], Youfang Lai[1,4], Huan Liu[1,2], Chenyang Shen[1,2], Yuncheng Zhong[1,2], Yiping Shao[2], Xinhua Li[3], Bob Liu[3], Xun Jia[1,2]

[1]innovative Technology Of Radiotherapy Computations and Hardware (iTORCH) Laboratory, Department of Radiation Oncology, University of Texas Southwestern Medical Center, Dallas, TX, 75235
[2]Department of Radiation Oncology, University of Texas Southwestern Medical Center, Dallas, TX, 75235
[3]Division of Diagnostic Imaging Physics, Department of Radiology, Massachusetts General Hospital, 55 Fruit Street, Boston, MA 02114
[4]Department of Physics, University of Texas Arlington, Arlington, TX, 76019
Emails: xun.jia@utsouthwestern.edu, KYANG11@mgh.harvard.edu

Motion management is a critical component of image guidance radiotherapy for lung cancer. We previously proposed a scheme using kV scattered x-ray photons for marker-less real-time image guidance in lung cancer radiotherapy. This study reports our recently progress using the photon counting detection technique to demonstrate potential feasibility of this method and using Monte Carlo (MC) simulations and ray-tracing calculations to characterize the performance. In our scheme, a thin slice of x-ray beam was directed to the target and we measured the outgoing scattered photons using a photon counting detector with a parallel-hole collimator to establish the correspondence between detector pixels and scatter positions. Image corrections of geometry, beam attenuation and scattering angle were performed to convert the raw image to the actual image of Compton attenuation coefficient. We set up a MC simulation system using an in-house developed GPU-based MC package modeling the image formation process. We also performed ray-tracing calculations to investigate the impacts of imaging system geometry on resulting image resolution. The experiment demonstrated feasibility of using a photon counting detector to measure scattered x-ray photons and generate the proposed scattered x-ray image. After correction, x-ray scattering image intensity and Compton scattering attenuation coefficient were linearly related, with $R^2$ greater than 0.9. Contrast to Noise Ratios of different objects were improved and the values in experimental results and MC simulation results agreed with each other. Ray-tracing calculations revealed the dependence of image resolution on imaging geometry. The image resolution increases with reduced source to object distance and increased collimator height. The study demonstrated potential feasibility of using scattered x-ray imaging as a real-time image guidance method in radiation therapy.

*The first two authors contributed equally.

## 1. INTRODUCTION

Image guided radiotherapy (IGRT) uses certain imaging modalities to visualize patient anatomy prior to or during treatment delivery to ensure treatment accuracy and hence reduces tumor targeting uncertainty and treatment planning margin sizes (Bujold *et al.*, 2012). Cone beam computed tomography (CBCT) is the most widely used method for pre-treatment patient positioning(Jaffray and Siewerdsen, 2000). 4D-CBCT has also been developed and increasingly used to provide respiratory phase resolved CBCT images for treatment sites that are affected by respiratory motions, such as lung and liver (Sonke *et al.*, 2005). Lately, magnetic resonance imaging (MRI) scanners have been integrated with medical linear accelerators (LINACs) to offer MRI-based image guidance with the advantages of high soft-tissue contrast and absence of ionizing radiation (Raaymakers *et al.*, 2017). However, for the tumors in lung and upper abdominal regions, pre-treatment images may not well represent patient anatomy during the treatment delivery due to notable changes in respiration amplitude, phase, and baseline (Keall *et al.*, 2006). Hence, there is a strong need for methods to derive real-time anatomy motion information to ensure treatment accuracy and patient safety.

Motion information may be obtained by tracking implanted fiducial markers or certain anatomical surrogates (Mao *et al.*, 2008; Cerviño *et al.*, 2009; Chi *et al.*, 2019; Ng *et al.*, 2012; Chi *et al.*, 2017). However, the maker-based methods introduce risks to patients due to the invasiveness of the procedure, and marker migration leads to the concern of tumor position accuracy. Surrogate-based methods encounter the problem of unreliable correlations between the surrogates and tumor motion. To directly image the anatomy in real time, the challenges are twofold. First, within a short time interval, it is difficult to measure a large amount of data that are sufficient to derive an image containing high-quality anatomical information. Second, it is also a challenge to perform data processing in real time, such as image reconstruction, to generate the image. At present, existing approaches may not fully address these two challenges. X-ray projection-based methods suffer from the problems of low image contrast and visibility of the target due to the projection nature that superimposes 3D anatomy onto a 2D plane (Lin *et al.*, 2009). Digital tomosynthesis is able to improve image resolution along the projection depth direction, by sacrificing temporal resolution (Godfrey *et al.*, 2006; Maurer *et al.*, 2008). Ultrasound-based imaging methods can provide images with a high temporal resolution, but their uses for IGRT are limited to only a certain number of scenarios due to interference of the probe to the therapeutic beam and possible distortion of the patient body surface (O'Shea *et al.*, 2016; Fontanarosa *et al.*, 2015). Recently, MR-LINAC allowed acquisitions of MR images during treatment delivery (Raaymakers *et al.*, 2017). Current technology for clinical use offers the imaging capability with <10 planar frames/second, and clinical adoption of this novel technology is on the rise.

Lately, it has been proposed to image patient anatomy and track tumor motion in real time by measuring scattered x-ray photons. X-ray scattering has been long considered as an enemy in x-ray based tomographic image modalities, as it reduces image contrast, leads to artifacts in the reconstructed images, and causing errors in image intensity (Jarry *et al.*, 2006; Zhu *et al.*, 2009). However, the scattered x-ray photons indeed carry anatomical information and if used properly, they may be employed to derive the image for image guidance purpose. In 2016, our group proposed a kV scattered x-ray imaging scheme that uses scattered kV x-ray photons for real-time imaging and motion monitoring in lung cancer radiotherapy (Yan *et al.*, 2016). By shining a thin slice of kV x-ray beam to the moving target region and measuring outgoing scattered photons, an image representing attenuation coefficient of Compton scattering in the image plane may be derived. Simulation studies were performed to investigate the feasibility of this approach, but experimental study was not possible at that time due to challenges in detecting scattered x-ray photons. In fact, the incoming photons are scattered to the entire $4\pi$ spherical angular space and are attenuated by the patient body, which makes the number of photons at the detector low. The required collimation in this technology further reduces the number of photons usable to generate a useful image. In 2018, Jones et al. and Redler et. al. proposed to measure MV scattered x-ray photons for imaging purpose (Jones *et al.*, 2018; Redler *et al.*, 2018). The high incident photon flux of a therapeutic MV x-ray beam increases number of scattered photons, making the detection possible. They have also recently demonstrated feasibility of tumor tracking in phantom studies (Jones *et al.*, 2020). Nonetheless, this method may encounter challenges in clinical applications caused by the use of a broad therapeutic beam, the occlusion of the incoming beam by the multi leaf collimator (MLC), and the suboptimal imaging geometry for tumor tracking purpose (see discussion section).

Recently, we made a breakthrough in our kV scattered x-ray imaging method. Utilizing advanced photon-counting detection (PCD) technology, we were able to demonstrate feasibility of this method in phantom studies under a realistic clinical setting. In this paper, we will report our experiment studies, as well as Monte Carlo (MC) simulation studies to validate the experiments. We will also report our computations in real patient cases to demonstrate the feasibility of tumor tracking and in a phantom case using ray-tracing method to investigate the impacts of geometric factors on resulting image resolution.

## 2. MATERIALS AND METHODS

### 2.1 Overview of the method

In the kV scattered x-ray imaging scheme, a narrow slit was used to collimate an incoming x-ray beam into a fan-beam form along the patient superior-inferior (SI) direction, as shown in Figure 1(a) (Yan *et al.*, 2016). Generally speaking, a tumor moves in the 3D space and a volumetric image is needed to completely describe the motion-induced anatomy variations. However, since the motion along the SI direction usually dominates, a 2D plane containing the SI direction may contain most information regarding the tumor motion. In the setup with the planner x-ray beam, the incoming x-ray photons were scattered off this plane and carry anatomy information in the plane. A parallel-hole collimator in front of the detector was used to impose a one-to-one correspondence between a detector pixel and a position on the imaging plane. Specifically, as shown in Figure 1(b), the x-ray photons were emitted from the source $S$ collimated in a planar form. After being scattered off this plane, photons detected at a detector pixel $D$ must come from the position $F$ of the plane because of the use of the collimator in front of the detector, or from the vicinity of $F$ due to the finite thickness of the incoming x-ray beam and the finite resolution imposed by the collimator. Note that although the Figure 1(b) illustrates the geometry with a scattering angle $\beta = 90°$, it is not necessary for the system to have this exact scattering angle. For instance, the set up in Figure 1(a) is possible. Once the system geometry is known, mapping from the measurements at the detector to the information in the imaged plane can be achieved based on the geometry.

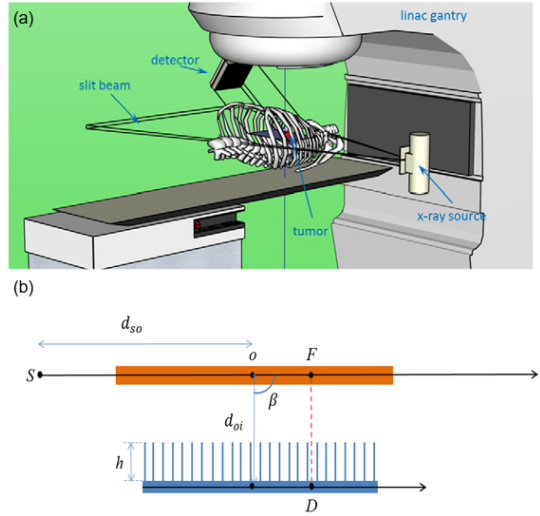

**Figure 1.** Top: Illustration of scattered x-ray imaging setup on a LINAC for tumor tracking. Figure reproduced from (Yan *et al.*, 2016) with permission. Bottom: A side-view illustration of the imaging principle of the system.

The first-order Compton scattered photons captured by the detector at the pixel $D$ can be written as

$$g(D) \sim f(F) \exp\left[-\int_S^F \mu(x)dl\right] \mu_C(F) \frac{d\sigma_C}{d\Omega}(F) \exp\left[-\int_F^D \mu(x)dl\right], \tag{1}$$

where $f(F)$ is the incoming x-ray intensity towards the voxel $F$. The two exponential functions describe x-ray attenuations along the incoming and outgoing ray lines and $\mu$ and $\mu_C$ are x-ray total attenuation coefficient and Compton linear attenuation coefficient, respectively. $\frac{d\sigma_C}{d\Omega}(F)$ is the differential Klein-Nishina (KN) cross section (Klein and Nishina, 1929) of the photon scattering angle at the voxel $F$. Based on this expression, the image $\mu_C(F)$ in the target region can be derived from the measured $g(D)$ and the known $f(F)$, $\frac{d\sigma_C}{d\Omega}(F)$, as well as the two exponential terms as:

$$\mu_C(F) \sim g(D) / \left\{ f(F) \exp\left[-\int_S^F \mu(x)dl\right] \frac{d\sigma_C}{d\Omega}(F) \exp\left[-\int_F^D \mu(x)dl\right] \right\}. \tag{2}$$

In this expression, only the first-order Compton scattering events were considered. Hence, $\mu_C(F)$ derived as such is an approximation, since the actual measurement $g(D)$ contains photons from first-order scattering events of other types and those from higher-order events. The validity of this approach for image guidance can only be demonstrated by subsequent experimental and computational studies.

## 2.2 Experiment setup

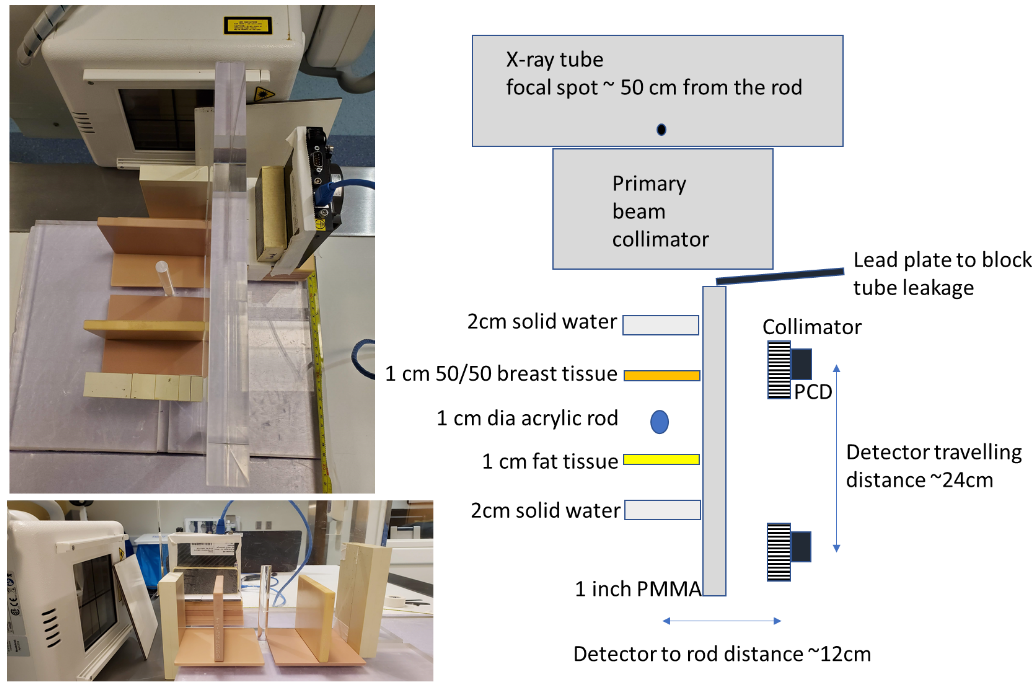

**Figure 2.** Experimental setup. Left: Top and side view of the setup. Right: Diagram showing setup details of the experiments.

The experimental measurements in this study were performed using a clinical radiographic system (DX-D600, AGFA, Morse, Belgium). The primary beam was collimated into a narrow slit (90mm x 5mm at 50 cm from the focal spot) by manually adjusting the collimator associated with the radiographic system. The x-ray technique was 120 kVp (half valued layer (HVL) of 4.9 mm Al), 200 mAs (500 mA, 400 ms), and a large focal spot. A Cadmium Telluride (CdTe, 0.75mm and 2.0mm) based photon counting detector (PCD) (XC-ACTAEON, Direct Conversion AB, Danderyd, Sweden) was used to detect scattered photons at about the 90-degree angle from the primary beam. The active area of the detector was 25.6 mm x 25.6 mm with pixel dimensions of 0.1 mm x 0.1 mm. With an energy threshold set at 10 keV, the scattered photons with energy above that threshold were counted at each pixel with a frame rate of 1 Hz, and the count arrays were saved into images. Without synchronization of the X-ray tube and the detector, a total of 10 frames (in 10 seconds) were acquired to ensure fully capture of each entire exposure. To spatially resolve the detected photons, a parallel-hole collimator, which was 25 mm in height, with 1 mm hole size and 0.1 mm lead septa, was placed at the front surface of the detector. The back side of the collimator was placed against the front surface of the PCD.

As shown in Figure 2, to simulate a simple but realistic scenario of lung tumor tracking, different phantom objects were placed along the pathway of the primary and scattered photons. A 1 cm diameter acrylic rod was used to simulate the targeted tumor and placed at 50 cm away from the x-ray focal spot. Two layers of 2 cm thick solid water plates were placed at 10 cm before and after the rod, to simulate the chest wall cage (averaging effect of muscle and ribs). Two additional layers of 1 cm thick breast tissue equivalent plates were placed at 5 cm before and after the rod, to generate more realistic tissue structures around the chest. A uniform 2.5 cm thick PMMA plate was placed between the imaging objects and the PCD, to simulate the attenuation of scattered photons. While the PMMA plate would also scatter the scattered photons, the secondary (and higher order) scattered photons were not considered when deriving the image $\mu_C(F)$ in Eq. (2). The PCD was placed about 12 cm from the rod target. To avoid the detection of x-ray tube leakage signal, a 5 mm solid lead plate was placed between the x-ray tube and the PCD. Due to its limited view size, the PCD was translated every 2 cm (1 cm around the rod) along the primary beam direction to provide a full coverage of all the imaging targets. The individually acquired images (each with 25.6 mm coverage and slight overlapping on the edges) were numerically merged together based on the corresponding location of the PCD in a postprocessing step to obtain a large image covering the entire region of interest.

## 2.3 Monte Carlo Simulation of experiments

We performed MC simulation studies to validate the experimental studies. The MC simulation system was based on an in-house developed GPU-based MC package gMCDRR for x-ray photon transport simulation of the kV energy range (Jia *et al.*, 2012). We designed a digital phantom containing objects in the same geometry locations and with the same material compositions as those in the experiments, as shown in Figure 3. A slit x-ray beam was directed to the acrylic rod. A poly-energetic x-ray spectrum of 120 kVp with HVL of 4.9 mm Al was generated using the SpekCalc tool (Poludniowski and Evans, 2007; Poludniowski, 2007) and was used in the simulation. A point x-ray source was set 50 cm from the rod. The incoming x-ray beam was collimated to a size of $0.5 \times 10$ cm$^2$ at the rod center. The imager was modelled as a $25 \times 10$ cm$^2$ array with pixel dimensions of $0.1 \times 0.1$ cm$^2$. It was placed at 12 cm away from the incoming x-ray beam. An ideal x-ray detector that captured all photons hitting it was used. A 2.5 cm parallel-hole collimator with the hole dimensions of $0.99 \times 0.99$ cm$^2$ and septa thickness of 0.01 cm was placed immediately in front of the detector. Scattered photons hitting on the collimator septa were assumed to be absorbed, and all the scattered photons hitting on the ideal detector were captured. The total number of source photon histories was $1.5 \times 10^{10}$ in the MC simulation of experiments, such that the computed contrast to noise ratio (CNR) matched that measured in the experiments.

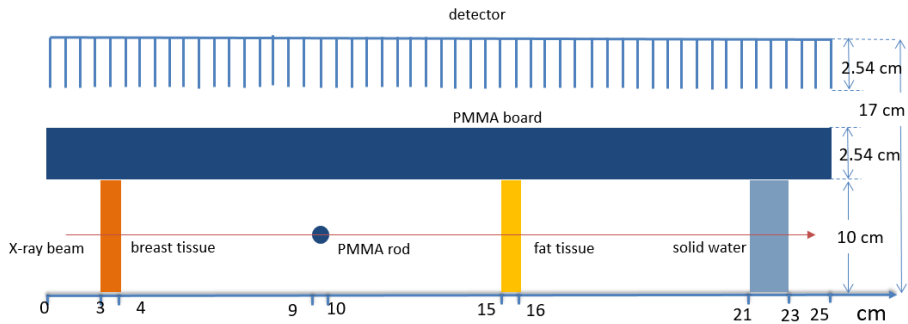

**Figure 3:** Setup of the MC simulation for the experiment case (geometry is not drawn to scale).

## 2.4 MC Simulation of patient cases

We also performed MC simulations for a patient case. We used a 4D-CT scan of a lung cancer patient to create the digital phantom. The correlation between CT numbers and tissue parameters in (Schneider *et al.*, 2000) was used during phantom creation. The size of the phantom was $42 \times 24 \times 20\ cm^3$ with voxel dimensions $0.1 \times 0.1 \times 0.1\ cm^3$. We considered the phase 0 (maximum exhale) and phase 5 (maximum inhale) in a respiratory cycle to study the anatomy change with tumors in two extreme positions. The tumor moved $6\ mm$ along the SI direction between these two respiratory phases. A slit x-ray beam collimated to the coronal plane was used. The size of the x-ray beam was $0.5 \times 10\ cm^2$ at the treatment isocenter of this case, which located approximately in the middle of the tumor. The focus spot was $100\ cm$ away from the tumor isocenter to follow the typical setup of an x-ray tube mounted on the LINAC. The x-ray spectrum was 120 kVp. A homogeneous photon fluence emitted by the x-ray source was used. An imager was placed in parallel to the x-ray plane but at $24\ cm$ away from the isocenter. The size of the detector was $42 \times 10\ cm^2$ and pixel dimensions of $0.1 \times 0.1\ cm^2$. The same parallel-hole collimator was used in the simulation. The total number of source particle histories was increased to $10^{11}$ in this case to obtain reasonable CNRs due to the higher x-ray attenuation caused by the tissues on the incoming and outgoing beam paths.

## 2.5 Impacts of imaging geometry on resolution

MC simulation can realistically model the image formation process. However, a large number of source photons have to simulated to obtain results with a low level of noise caused by statistical fluctuation. To investigate the impacts of various factors of the imaging system, e.g. collimator height and location, on the resulting image quality, MC simulation may not be appropriate due to high computational burden. Hence, we also performed computations using a simplified analytical approach that only considered the first order Compton scatter effect via ray-tracing calculations. As such, the system setup was the same as that of the MC simulations. To investigate the dependence of image resolution on system geometry, we built a digital phantom with a size of $42 \times 10\ cm^2$ and pixel dimensions of $0.1 \times 0.1\ cm^2$. A small spherical shape 'tumor' object with the material of soft tissue and density of $\rho = 1.0$ g/cm³ was placed in the center of the phantom, which located at the isocenter. The diameter of the sphere was 0.6 cm. The rest of the phantom space was filled by a background material of lung tissue with a density of $\rho = 0.3$ g/cm³. The incoming x-ray beam was directed to the tumor. The signal at the pixel $D$ of the imager can be written as

$$g(D) = \sum_F f(F) \int_0^{E_p} \eta(E)\, e^{-\int_S^F \mu(x,E)\mathrm{d}l} \mu_C(F,E) \frac{\mathrm{d}\sigma_C}{\mathrm{d}\Omega}(F) e^{-\int_F^D \mu(x,E)\mathrm{d}l} \mathrm{d}E, \qquad (3)$$

where the summation notation $\Sigma_F$ here represents the summation over all the possible voxels $F$ such that the x-ray path from $S$ scattered at $F$ can reach the area of the pixel $D$. $f(F)$ is the incoming x-ray intensity toward the voxel $F$ of the phantom, $\eta(E)$ is the energy distribution function of the x-ray spectrum, and $E_p = 120$ keV is the upper limit of the x-ray energy. $\mu(x, E)$ and $\mu_C(F, E)$ are the total and Compton attenuation coefficients of corresponding voxels and energy, and $\frac{d\sigma_C}{d\Omega}(F)$ is the differential KN cross section (Klein and Nishina 1929) of the photon scattering angle at voxel $F$. The line integrals in the exponential functions were performed using the classical Siddon's ray-tracing algorithm (Siddon, 1985) for calculating the radiological path in the 3D space. Extensive computations with different detector to isocenter distance $d_{oi}$ ranging between 24 cm to 48 cm and collimator height $h$ between 2.54 cm to 5.08 cm were performed.

## 2.6 Data processing and evaluations

Postprocessing calculations were needed to convert the data at the detector $g(D)$ into the image of actual image $\mu_C(F)$ based on Eq. (2). We assumed that the incoming photon fluence, $f(F)$ was homogeneous, as the intensity variation over the illumination area was relatively small. The KN differential cross section (Klein and Nishina, 1929) can be written as $\frac{d\sigma_C}{d\Omega}(F) \propto P^2[P + \frac{1}{P} - 1 + \cos^2\theta]$, where $\theta$ is the Compton scattering angle at voxel $F$，and the ratio of photon energy after and before the scattering $P = 1/(1 + (1 - \cos\theta) \cdot (\frac{E}{m_e c^2}))$, with $m_e c^2 = 511$ keV being the electron rest energy and $E$ being the initial photon energy. We computed this term using the mean energy $\bar{E}$ of the poly-energetic x-ray beam as an approximation to $E$. The incoming and outgoing attenuation terms $\exp[-\int_S^F \mu(x)dl]$ and $\exp[-\int_F^D \mu(x)dl]$ were computed using the Siddon's algorithm. The total attenuation $\mu(x)$ here was also computed at the mean energy $\bar{E}$. After dividing the raw image $g(D)$ by these correction terms, we obtained the actual image $\mu_C(F)$ that represents the anatomy in the plane of interest.

On additional step processing the experimental data was the removal of a grid pattern caused by the parallel hole collimator. The use of the collimator created a grid pattern on the raw image $g(D)$, see for example Figure 4(a). Before applying the abovementioned postprocessing steps, we used a low-pass filter to remove this pattern. The filtering process was realized using Fourier Transform.

For both the raw and corrected images of the experimental case and the patient case in MC simulations, we evaluated the image quality quantitatively by contrast-noise-ratio (CNR). The CNR is defined as $\text{CNR} = |A(f^F) - A(f^B)|/S(f^B)$, where $A(.)$ standards for the average over the regions, $S(.)$ represents the standard deviation of the background, $f^F$ is the acrylic rod or the tumor (foreground) and $f^B$ is the nearby region (background).

In the last study step using ray-tracing calculations to investigate the impacts of geometry configuration on image resolution, for the image obtained in a specific setup, we deconvolved the image using the known ground truth image that contained the tumor of diameter 0.6 cm to obtain the blurring kernel. We then measured the full width at half maximum (FWHM) of the blurring kernel, which was used as the metric to measure image resolution. The dependence of the FWHMs as geometry configurations was investigated.

## 3. RESULTS

### 3.1. Experimental results

Figure 4(a) shows one example raw image acquired from PCD corresponding to the acrylic rod. Although the grid structure of the collimator was clearly visible, it did not affect the visualization of the object boundary. The mean photon counts per pixel was ~31 within the selected ROI. We would like to remind the readers about the small pixel size of 0.1 mm x 0.1 mm. We summed over photon counts to reduce the resolution to ~1mm and further enhance the signal to noise ratio of the raw data. Figure 4(b) shows the processed image after removing the collimator pattern by applying a low-pass filter. Figure 4(c) and 4(d) shows the combined raw image and that after applying the low-pass filter from the total of 14 different PCD locations. The raw image can accurate depict different objects and their relative locations, although the intensity generally decayed along the beam direction due to attenuation of the primary photon beam.

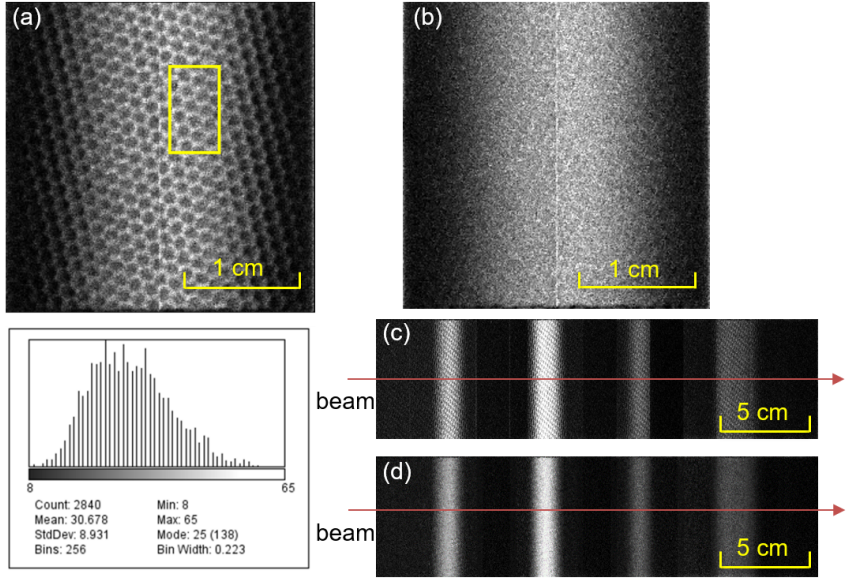

**Figure 4:** Experimental raw image. (a) An example raw image acquired from PCD. (b) The processed image after removing the collimator pattern using a low-pass filter. (c) and (d) The combined raw image from the total of 14 different PCD locations and that after the low-pass filter.

### 3.2. Simulation of experiments

Figure 5(a) and (b) show the comparison of images from experimental measurement (after applying the low-pass filter) and MC simulation before the postprocessing steps. $g_i$ and $g_i'$ represent the raw images of experiment and simulation with subscript $i = 1, 2, 3$ and $4$ representing breast tissue, acrylic rod, fat tissue and solid water, respectively. Photon numbers were set to $1.5 \times 10^{10}$ in the MC simulation, to ensure the noise level of MC simulation results are the same as that of experimental results. Visually, an agreement between simulation and experiment results was achieved.

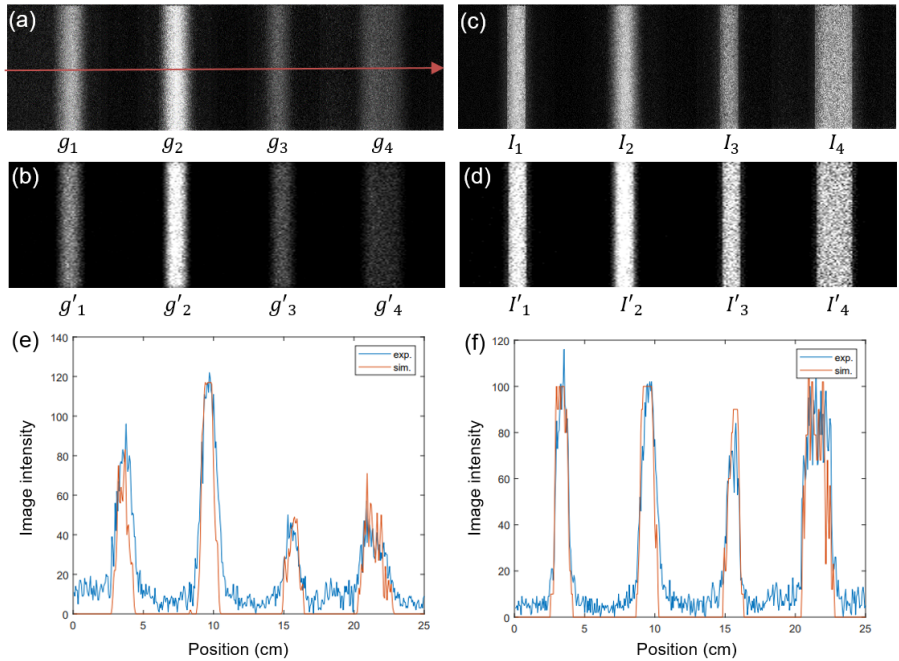

**Figure 5:** (a) Raw and (c) corrected images obtained from the experiments. (b) Raw and (d) corrected images obtained from the MC simulations. (e) and (f) are comparisons between image intensity profiles in experiments and MC simulations along the line in (a).

The correction process described in Sec 2.5 was performed to obtain the corrected images, as shown in Figure 5(c) and 5(d), where $I_i$ and $I'_i$ are corrected images of Compton attenuation coefficient $\mu_C$. Figure 5(e) and 5(f) are the comparison of image intensity profiles between experimental and MC simulation results before and after the postprocessing steps. It is expected that the image value $I_i$ is correlated with $\mu_C$ under the mean energy of the beam. Figure 6 shows the linear fit with the R-squared value $R^2 = 0.91$, which indicates the feasibility of using the corrected image values to represent different tissues.

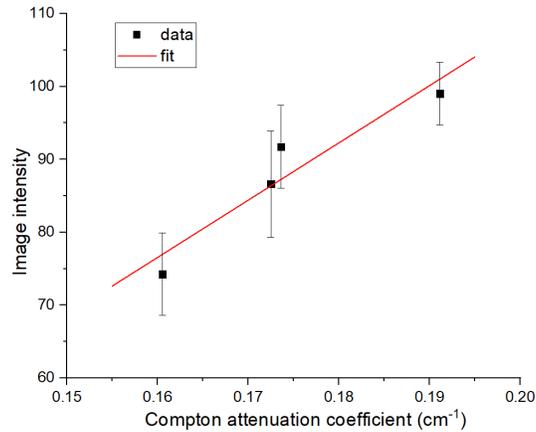

**Figure 6:** Linear correlation between corrected image intensities and Compton attenuation coefficients.

CNR of various objects in the experiment and simulation studies were evaluated and the results are listed in Table 1. The CNR values in the experiments and MC simulations matched each other. Moreover, the CNR of the processed images were consistently higher than those of the raw images, because the correction step improved image contrast.

**Table 1.** CNR of various objects in measurement and simulation of experiment.

|  | $g_1$ | $g_2$ | $g_3$ | $g_4$ | $I_1$ | $I_2$ | $I_3$ | $I_4$ |
|---|---|---|---|---|---|---|---|---|
| Experiment | 8.9 | 15 | 5.2 | 4.6 | 18 | 19 | 14 | 17 |
| MC Simulation | 9.5 | 14 | 5.3 | 4.7 | 18 | 15 | 17 | 18 |

### 3.3. Simulation of patient cases

In the MC simulation of the patient case, the total photon number was set to be $10^{11}$. Figure 7(a) shows the 4D-CT images of the patient at two respiratory phases used in the simulation and there was a tumor motion of 6 mm in the SI direction between the two phases. Figure 7 (b) and (c) are raw and corrected images obtained via MC simulations, respectively. In the raw images, the image intensity gradually decayed along the beam direction due to x-ray attenuation. A high intensity region inside the lung can be observed corresponding to the tumor locations. After correction, the tumor image can be visualized more clearly. CNRs for the raw and corrected simulation images were 2.7 and 4.9. Here the foreground was the tumor and the background was the lung tissue around the tumor.

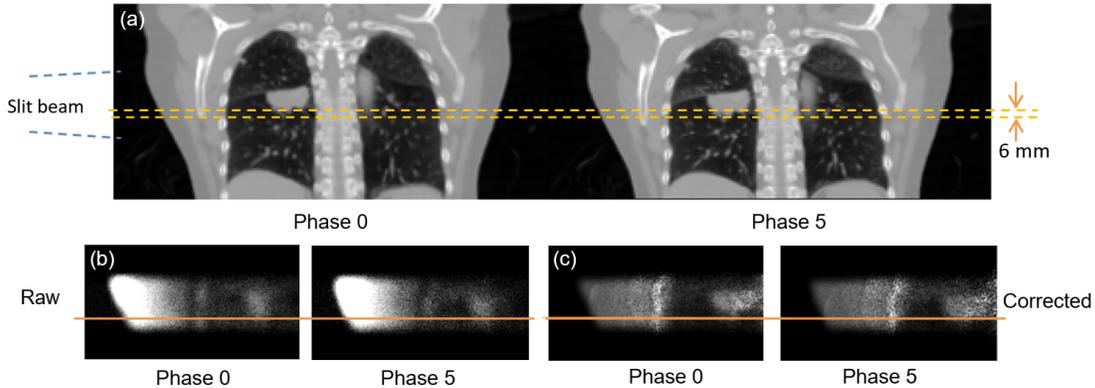

**Figure 7:** (a) 4D-CT scan of a patient used to create the digital phantom in the MC simulation. Images at two respiratory phases are shown. (b) and (c) are raw and corrected images obtained via MC simulations, respectively.

### 3.4. Impacts of imaging geometry on resolution

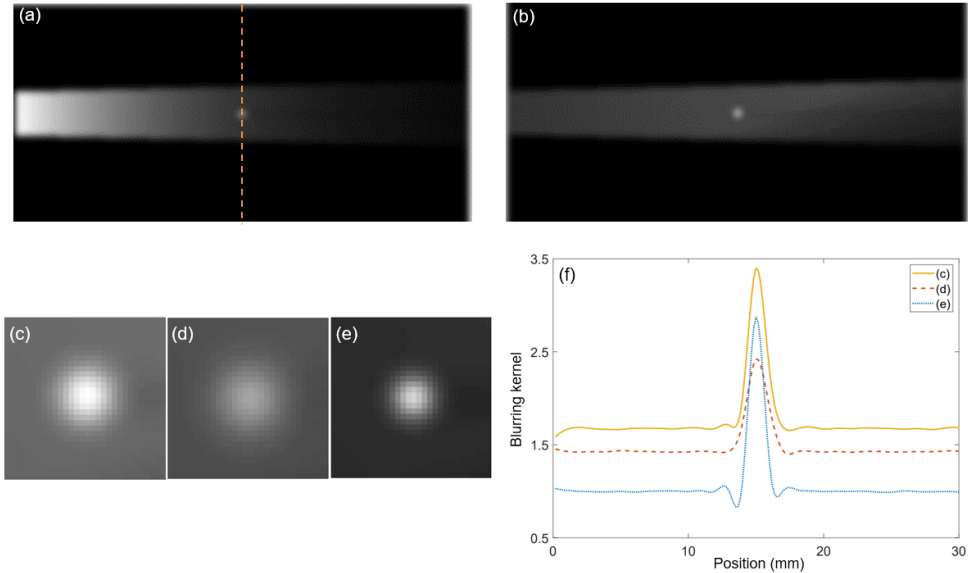

**Figure 8:** (a) Raw image of ray-tracing calculation for the phantom in a representative setup of $d_{oi} = 24$ cm, $h = 2.54$ cm. (b) Corrected image corresponding to (a). (c)-(e) are corrected images for $d_{oi} = 24$ cm, $h = 2.54$ cm, $d_{oi} = 48$ cm, $h = 2.54$ cm, and $d_{oi} = 24$ cm, $h = 5.08$ cm cropped around the object of interest. (f) Comparison of blurring kernels in the three cases.

We performed computations to investigate the impacts of imaging geometry on resulting image resolution. Specifically, we enumerated simulations for different distances between the imager and the object $d_{oi}$ ranging in 24 cm to 48 cm and collimator height $h$ ranging in 2.54 cm to 5.08 cm. Figure 8 shows the results in three representative cases. When the imager-to-object distance was increased, the resulting image became more blurred, because the region in which a detector pixel can receive scattered photon became larger. The same effect happened when the collimator height was reduced. Meanwhile, as the detector was moved away from the object, the overall image intensity was reduced, because of reduced photon numbers reaching the detector. After deconvolving the resulting image based on the known ground truth image containing the tumor with the diameter of 0.6 cm, we obtained the blurring kernels under different setups, as shown in Figure 8 (c)-(e) as two-dimensional image and in Figure 8(f) as one-dimensional profiles. The FWHMs of the blurring kernels are shown in Figure 9. The maximal FWHM was 1.98 mm, which occurred in the case with the longest imager-to-object distance and shortest collimator height. In contrast, the minimal FWHM was 0.98 mm, achieved in the setup with opposite parameters.

## 4. DISCUSSIONS

This study demonstrated the potential feasibility of using the photon counting detection technique to realize the scattered x-ray imaging method. However, there are a few challenges remaining for future implementation. Due to the finite size of the primary collimation and limited rejection power of the collimator, there is unavoidable blurring of the object edge. This might affect the tracking of very small sized tumor. Based on the calculated FWHM about 1-2 mm, our expectation is to be able to track tumors with size greater than 1 cm.

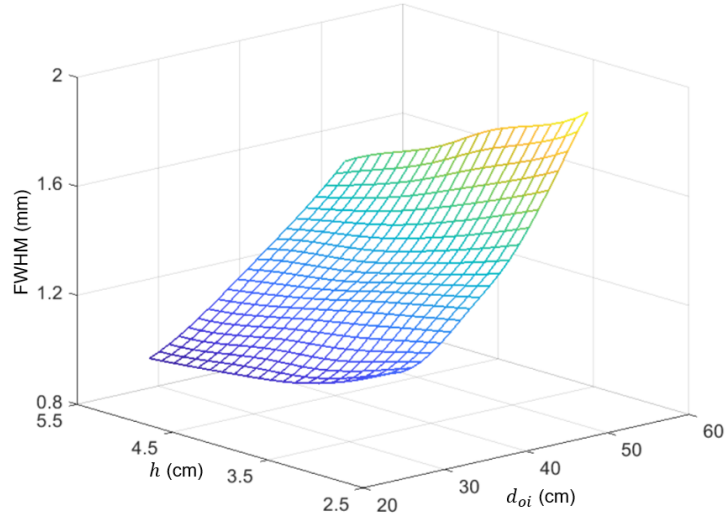

**Figure 9:** FWHM of the blurring kernels as a function imager-to-object distance $d_{oi}$ and collimator height $h$.

In the current set up in experiments, the x-ray focal spot is about 50 cm from the targeted rod and the PCD is about 12 cm from the rod. This indicates that both the tube and the detector will be very close to the patient body (probably 5 cm from the skin). This might be a limitation for clinical implementation, given all the collision constraints around the treatment bed. For the tube side, possible solution is to increase the distance of tube to isocenter, as shown in our MC simulations. Increasing tube output or relaxing the primary collimation would be needed to increase incoming photon counts of the primary beam. For the detector side, it might need the increase of the grid ratio of the collimator, without losing too much detected signal. In addition, advanced image processing techniques, such as deep learning based approaches (Shen *et al.*, 2020), may be applicable to help improving image signal and facilitate the task of tumor tracking.

Based on the previous novel studies (Jones *et al.*, 2018; Redler *et al.*, 2018) and the current study, scattered x-ray photons in both the MV therapeutic beam and the kV imaging beam may be used to derive images for real-time image guidance purpose. We think the use of the kV beam has certain advantages. First, the MV beam is used for therapeutic purpose. While it is a novel approach to use it simultaneously for imaging purpose, the imaging function would be affected by beam settings required by therapy. For instance, a broad beam is often delivered towards the tumor, and hence the measured scattered signal is the integration along the direction of the scattered photon direction. As illustrated in Figure 10, the measurement integrates scattered photons from all x positions. This leads to the lack of capability resolving image information along the x direction and possibly reduces image contrast due to superposition. In addition, the use of a multi-leaf collimator for intensity modulation in intensity modulated radiation therapy would collimate the therapeutic beam into irregular shapes, limiting the clinical uses of this beam to image the tumor. When the multi-leaf collimator blocks tumor, the detected tumor intensity is expected to be reduced. Second, a kV beam has its source located at 90-degree position relative to the MV beam on a standard

LINAC. Using the kV beam source provides useful information for tumor tracking. Due to the 2D measurement nature, measuring the scattered photons from either the MV or the kV beam can at most provide imaging information along two out the three principle directions. For tumor tracking purpose, it is critical to obtain information about tumor motions perpendicular to the therapeutic beam i.e. x, y directions in Figure 10, as the tumor may move out of the beam, leading to substantial under coverage. The motion along the third direction, z is less important, as its impact on delivered dose is characterized by the depth dose of the therapy beam, which is much more gradual. In the case using a kV beam for imaging, the scattered photons can derive motion in the x-y plane. In contrast, the measurement of scattered photons from the MV beam can only provide information in the y-z plane, which is less optimal to provide clinical useful tumor motion information. However, it is also noted that the use of the kV beam would introduce additional ionizing radiation to the patient and the signal is relatively weak. It is an ongoing study to overcome these limitations of the kV beam approach aiming at bringing this technology to clinical practice.

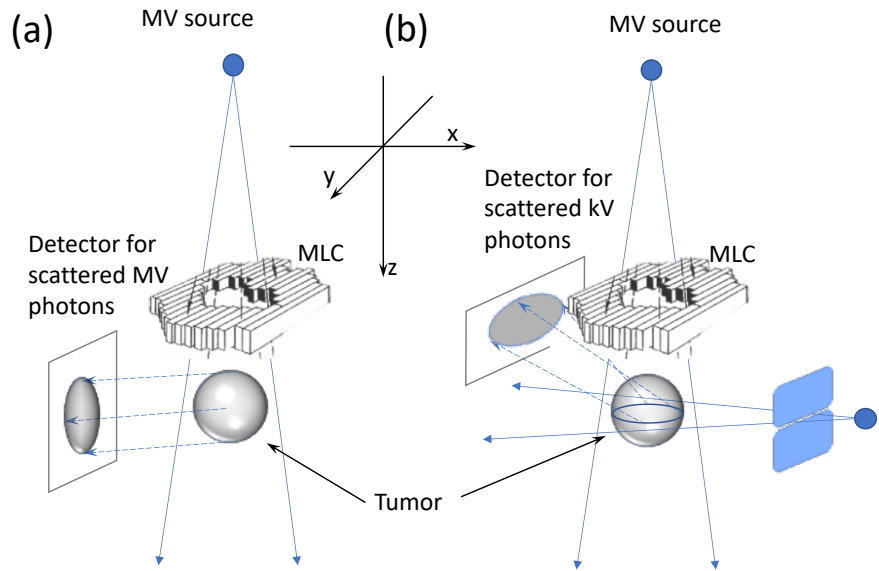

**Figure 10.** Comparison of imaging geometry in MV and kV based scattered photon imaging.

## 5. CONCLUSION

In this study, we investigated a previously proposed scheme using kV scattered x-ray photons for marker-less real-time image guidance in lung cancer radiotherapy. In this scheme, a thin slice of x-ray beam was directed to the target and we measured the outgoing scattered photons with a parallel hole collimator to establish the correspondence between pixels and scatter positions. Image corrections of geometry, beam attenuation and scattering angle were performed to convert the raw image to the actual image representing attenuation coefficient of Compton scatter effect. Employing the photon counting detection technique, we were able to measure scattered photons and used it to form the resulting image. We performed MC simulations and ray-tracing calculations to characterize the performance the imaging system. It was found that the corrected image intensity and Compton scattering attenuation coefficient were linearly correlated ($R^2 = 0.91$). Contrast to Noise Ratios of different objects were improved by the correction step and the values in experimental results and MC simulation results agreed with each other. Ray-tracing calculations revealed the dependence of image resolution on imaging geometry. The image resolution increases with reduced imager to object distance and increased collimator height. The study demonstrated potential feasibility of using scattered x-ray imaging as a real-time image guidance method in radiation therapy.

*Acknowledgement* This study is supported in part by grants from National Institutes of Health (R01CA214639, R01CA227289, R01CA237269, R01EB019438, and R01CA218402). The photon counting detectors were generously provided by Direct Conversion AB, Danderyd, Sweden. The authors would like to thank Dr. York Haemisch and Roland Neal for their support with the detectors.